# State-of-the Art Motion Estimation in the Context of 3D TV

**ABSTRACT**
Progress in image sensors and computation power has fueled studies to improve acquisition, processing, and analysis of 3D streams along with 3D scenes/objects reconstruction. The role of motion compensation/motion estimation (MCME) in 3D TV from end-to-end user is investigated in this chapter. Motion vectors (MVs) are closely related to the concept of disparities and they can help improving dynamic scene acquisition, content creation, 2D to 3D conversion, compression coding, decompression/decoding, scene rendering, error concealment, virtual/augmented reality handling, intelligent content retrieval and displaying. Although there are different 3D shape extraction methods, this text focuses mostly on shape-from-motion (SfM) techniques due to their relevance to 3D TV. SfM extraction can restore 3D shape information from a single camera data.

## A.1 INTRODUCTION

Technological convergence has been prompting changes in 3D image rendering together with communication paradigms. It implies interaction with other areas, such as games, that are designed for both TV and the Internet.

Obtaining and creating perspective time varying scenes are essential for 3D TV growth and involve knowledge from multidisciplinary areas such as image processing, computer graphics (CG), physics, computer vision, game design, and behavioral sciences (Javidi & Okano, 2002). 3D video refers to previously recorded sequences. 3D TV, on the other hand, comprises acquirement, coding, transmission, reception, decoding, error concealment (EC), and reproduction of streaming video.

This chapter sheds some light on the importance of motion compensation and motion estimation (MCME) for an end-to-end 3D TV system, since motion information can help dealing with the huge amount of data involved in acquiring, handing out, exploring, modifying, and reconstructing 3D entities present in video streams. Notwithstanding the existence of many methods to render 3D objects, this text is concerned with shape-from-motion (SfM) techniques.

Applying motion vectors (MVs) to an image to create the next image is called motion compensation (MC). This text will use the term "frame" for a scene snapshot at a given time instant regardless of the fact that it is 2D or 3D video. Motion estimation (ME) explores previous and/or future frames to identify unchanged blocks. The combination of ME and MC is a key part of video compression as used by MPEG 1, 2 and 4 in addition to many other video codecs.

Human beings get 3D data from several cues via parallax. In binocular parallax each eye captures its view of the same object. In motion parallax, different views of an object are obtained as a consequence of head shift. Multi-view video (MVV) refers to a set of *N* temporally synchronized video streams coming from cameras that capture the same real world scenery from different viewpoints and it is widely used in various 3D TV and free-viewpoint video (FVV) systems. The stereo video (*N = 2* videos) is a special case of MVV. Some issues regarding 3D TV that need further developments to turn this technology mainstream are

a) availability of a broad range of 3D content;
b) suitable distribution mechanisms;
c) adequate transmission strategies;
d) satisfactory computer processing capacity;
e) appropriate displays;
f) proper technology prices for customers; and
g) 2D to 3D conversion allowing for popular video material to be seen on a 3D display.

Video tracking is an aid to film post-production, surveillance and estimation of spatial coordinates. Information is gathered by a camera, combined with the result from the analysis of a large set of 2D trajectories of prominent image features and it is used to animate virtual characters from the tracked motion of real characters. The majority of motion-capture systems rely on a set of markers affixed to an actor's body to approximate their displacements. Next, the motion of the makers is mapped onto characters generated by CG (Deng, Jiang, Liu, & Wang, 2008).

3D video can be generated from a 2D sequence and its related depth map by means of depth image-based rendering (DIBR). As a result, the conversion of 2D to 3D video is feasible if the depth information can be inferred from the original 2D sequence (Fehn, 2006; Kauff, et al., 2007).

Light field refers to radiance as a function of location and direction of areas without occlusion. The final objective is to get a time-varying light field that hits a surface and is reflected with negligible delay. An example of a feasible dense light field aquisition system is one using optical fibers with a high-definition camera to obtain multiple views promptly (Javidi & Okano, 2002). For the most part, light field cameras allow interactive navigation and manipulation of video and per-pixel depth maps to improve the results of light field rendering.

Video restoration (denoising) is crucial to 3D TV and it continues to call for research because of difficulties such as nonstationarities and discontinuities in spatio-temporal signals as well as the need to minimize the error between the unknown original sequence and the restored one. Denoising can benefit from ME to manage large displacements due to camera motion.

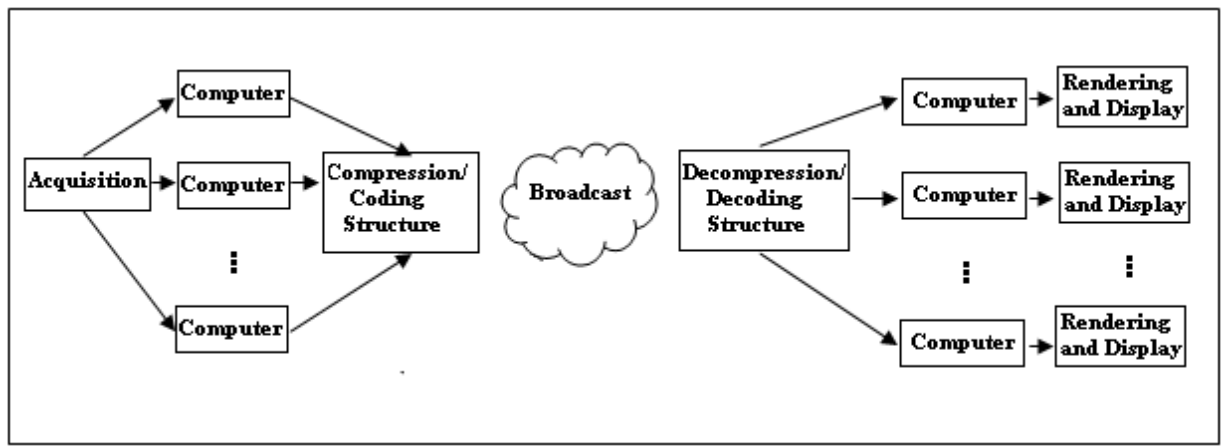

*Figure A.1 - End-to-end 3D TV System*

At least two video streams captured by different cameras synchronized in time are necessary to show 3D content. A 3D TV system should offer an adequate amount of viewpoints and this trait is called FVV by the MPEG Ad-Hoc Group on 3D Audio and Video (3DAV) (Smolic & McCutchen, 2004). A high number of pixels per viewpoint also helps rendering 3D scenarios. Current developments in hardware point towards a completely distributed design for acquisition, compression, transmission, and representation of dynamic scenes as well as total scalability when it comes to the implementation of all stages.

Figure A.1 illustrates a typical 3D TV system. The acquisition step has arrangements of synchronized cameras connected to the computers (producers) that generate/assemble the 3D video content. The producers combine several image sources from live, uncompressed streams and then, code them. Next, the compressed dynamic scenes are sent to distinct channels with the help of a broadcast network. Individual video streams are decompressed by decoders at the receiver side. The decoders are networked to a cluster of consumer computers in charge of rendering the proper views and direct them to a 3D display. A dedicated controller may transmit virtual view parameters to decoders and consumers. It can also be connected to cameras located in the viewing region for automatic display calibration. Scalability in the amount of captured, broadcast, and presented views can be achieved if a completely distributed processing is used.

A possible system realization applies a one-to-one mapping of cameras to projectors, provided images are rectified (camera calibration) to correct aligned. Because cameras must be uniformly spaced, this mapping lacks flexibility and it fails if the number of cameras differs from the amount of projectors. However, it scales fine. Another flexible scheme is an image-based rendering to synthesize views at the correct virtual camera positions where the geometric proxy for the scene is a single plane arbitrarily set.

Lately, academia and industries have paid a lot of attention to multimedia communications over networks. Video applications can be conversational (video telephony or teleconference) and streaming. In talkative applications, a peer-to-peer connection is created and both sides execute identical tasks, such as coding and decoding. Video streaming generally employs client-server architecture. The client queries the server for a particular video stream, and then a specific quantity of information is reserved. Subsequently, the server sends video so that the client is able to decode and display dynamic scenes in real-time. If packets are lost, the preliminary lag caused by the pre-rolling process typically permits the client to request some retransmissions; nevertheless a remaining rate of packet losses persists. Error concealment (EC) algorithms encompassing spatial, temporal, and spatiotemporal interpolation soothe damages caused by losses as seen in (Belfiore, Grangetto, Magli, & Olmo, 2003; Coelho, Estrela, & de Assis, 2009; Coelho, & Estrela, 2012b).

This chapter is organized as follows: section A.2 provides the necessary background on MCME; section A.3 explains how MCME aid all stages of an end-to-end 3D TV. Recommendations for future research are suggested in a separate section, which is followed by a final one compiling the major conclusions on MCME in the context of 3D TV.

## A.2 BACKGROUND AND DEFINITIONS
### A.2.1 Motion Compensation/Motion Estimation (MCME)

Motion analysis is necessary to understand visual surrounds as well as to improve coding, vision, object segmentation, tracking, and so on. In addition, MCME and depth data are a strongly related in 3D TV.

Tracking significant objects in video is the basis of many applications like video production, remote surveillance, robotics, interactive immersive games and comprehension of large video datasets helps to trim down the work required for a task and thus facilitating the development of 3D TV systems

Motion detection (MD) is the ability of a system to perceive motion and gather important occurrences. An MD algorithm senses motions, prompts the surveillance system to start acquiring a scene and it can evaluate the motion to take another course of action. So far, the intricacy of MV detection does not permit real-time realizations of this procedure. MEMC schemes based on the H.264 standard use a fixed block-size matching procedure. The motion of an object is directly proportional to its distance from the camera. Thus, the depth map is approximated by a constant associated to the estimated motion. Regrettably, this only holds for a small fraction of videos with camera panning across a stationary scene or with a still camera.

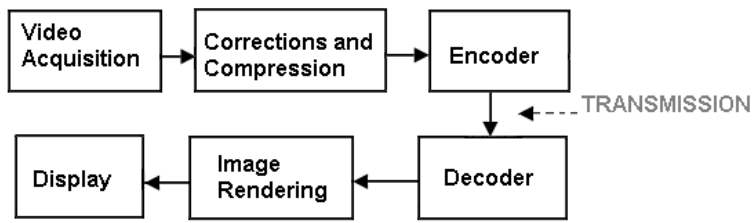

*Figure A2 – Simplified view of an end-to-end 3D TV system.*

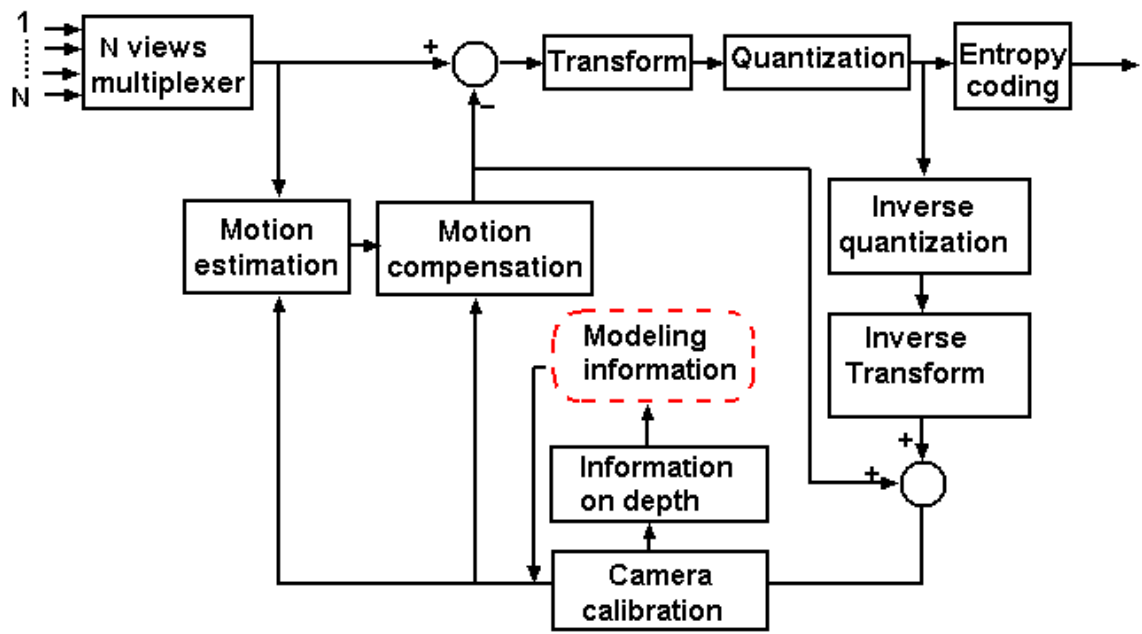

*Figure A.3 - A general 3D video coding architecture.*

MVs may be related to global motion or to local motion of specific regions. The MVs may be represented by: 1) a translational model only; 2) models that can approximate the motion of a real camera; 3) feature-oriented methods such as the Harris corner detector so as to locate a set of points that work as a signature of an object; and 4) feature-matching methods that find all corresponding features between frames. The most popular classes of motion estimation algorithms are described in the next 3 paragraphs.

The consistency of local grey-level patterns in successive images is an important assumption for optical flow (OF) estimation. Redundancy between adjoining frames can be used for compression where a frame is chosen as a reference and the next frames are predicted from this reference by means of ME (interframe coding). So, each frame can be predicted by considering the previous frame and the estimated motion. Optical Flow (OF) is the perceptible displacement of objects and special features in a scene caused by dislocation of image intensities (pixel grey levels) in a scene. There is a close relationship between ME, and stereo disparity measurement (Estrela, & Galatsanos, 2000; Garbas et al., 2006; Coelho & Estrela, 2012b). Lots of advances in ME and video compression have resulted from OF research, because it yields a dense motion field, and it helps estimating the 3D structure of its constituents. 3D Optical Flow (3D OF) is more sensible when it comes to motion awareness and mental maps generation of the environment. A typical approach for 3D OF is to compute time-varying depth maps using stereo algorithms, followed by the computation of range or scene flow from data. In practice, stereo is used to obtain depth maps and compute range flow. An interesting application of stereo is 3D scene generation for CG and 3D tracking (Wedel et al. 2008). The

A Block Matching Algorithm (BMA) locates matching blocks between video frames in order to estimate motion, that is, it picks up MVs compliant with the best correlation between pairs of blocks contained in successive frames. Hence, BMA takes advantage of the concept of temporal redundancy in the video sequence, which aims at increasing the effectiveness of interframe video compression and it has been adopted in MPEG-2 motion estimation (ME) due to its simplicity and effectiveness (Furht, 2008). However, the lack of contrast, noise and illumination changes can impair ME leading to an inverse problem. Once the MVs are found, they can be used to describe the transformation from one image to another. Algorithms for implementing ME are carried out by the codec.

Phase Correlation (PC) uses the Fourier Transform (FT) of adjacent or similar images to estimate the relative offset between them. PC is more robust to rotation and deformation of objects than other methods which turn it easier the identification of frequency signatures related to MVs. The FT represents only the inter-frame changes in terms of frequency distribution; hence there exists less data to be processed. Each motion corresponds to a peak in the 3D frequency spectrum given by the FT. So, the computer has just to track the peaks and assign the right MVs to them. Once the dominant motion has been estimated, then filtering along the motion direction can be performed (Lucchese, Doretto, & Cortelazzo, (2002))

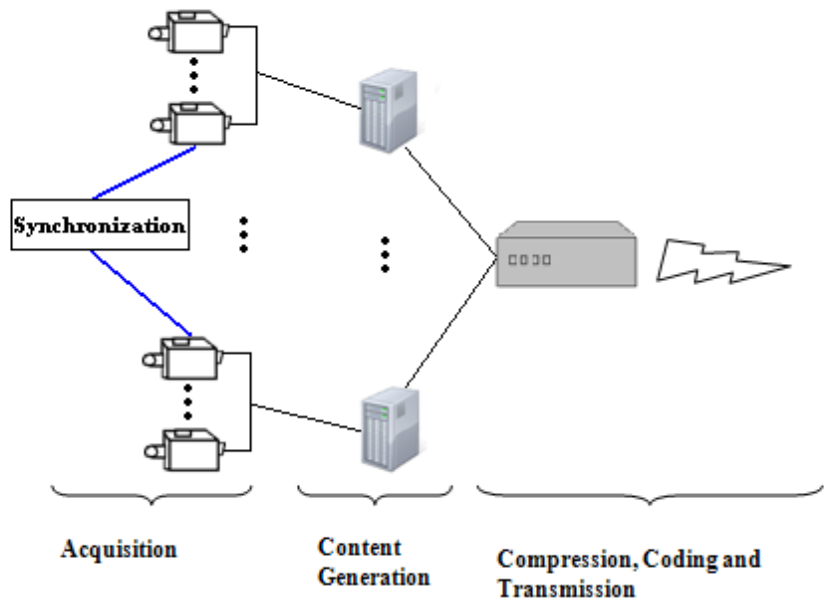

*Figure A.4 – Stages from the 3D TV producer.*

High compression and high coding efficiencies result from taking advantage of both spatial and temporal redundancies. A coder configuration using the closed-loop ME is called as a predictive video encoder.

H.264/MPEG-4 Part 10 or AVC (Advanced Video Coding) is an extremely popular standard for video compression, and for high definition sequences. This standard has a codec that relies on a BMA MC.

Scalable Video Coding (SVC) allows bitstreams that contain sub-bitstreams, where both conform to the standard. The base layer can be decoded by an H.264/AVC that does not support SVC. A subset bitstream can represent a lower spatial resolution, or a lower temporal resolution, or a lower quality video signal compared to the bitstream it is derived from. There are four possible types of scalability:

1) Temporal (frame rate) scalability structures the MC dependencies so that complete frames can be removed from the bitstream.
2) Spatial (frame size) scalability is related to the fact that dynamic sequences are coded at various spatial resolutions. The information and decoded samples from lower resolutions can be used to predict data or samples of higher resolutions aiming at decreasing the bit rate necessary to code the higher resolutions.
3) SNR/Quality/Fidelity scalability is concerned with coding at a particular spatial resolution but with different image qualities. The information and decoded samples corresponding to lower qualities can be used to estimate data or samples from higher qualities to lessen the bit rate used to code the higher qualities.
4) All scalability defined above can be combined

MVV Coding (MVC) permits the creation of bitstreams corresponding to several views of a scene which facilitates the implementation of stereoscopic 3D video coding. The Multi-view High Profile supports multiple views, and the Stereo High Profile is dedicated to stereoscopic video (two views).

Broadcasting problems can be alleviated by means of efficient coding and compression strategies. 3D TV broadcasting calls for the transmission of all views to multiple clients at the same time. The quantization can be described as a mapping of a continuous set of values (or a very large set of possible discrete values) to a relatively small discrete and finite set. The resulting quantized signal should be possible to represent with fewer bits than the original since the range of values is smaller. The process is lossy (*i.e.* not reversible). Many transforms have been proposed to trim down redundancies and the amount of data to be coded, such as the Principal Component analysis (PCA), Singular Value Decomposition (SVD), the Discrete Wavelet Transform (DWT), and the discrete cosine transform (DCT). Since most compression techniques are lossy, compression-induced distortion metrics are needed to appraise video coding algorithms. In practice, MSE and PSNR have been extensively used due to their clear physical meanings and simplicity, but they are far from portraying perception as the HVS.

The DCT is omnipresent in all video compression standards because it is well fitted with the BMA structure. Non-zero DCT coefficients of a translating area (global, constant-velocity, translational motion) are situated on a folded plane in the spatio-temporal frequency space and this fact helps to estimate motion in a 3D block of pixels.

Most visual coding standards (MPEG-1, MPEG-2, MPEG-4 Visual, H.261, H.263 and H.264) are based on predictive DPCM/DCT design (Hewage, Karim, Worrall, Dogan, & Kondoz, 2007; Richardson & al., 2003; Schwarz, Marpe, & Wiegand, 2006). An MPEG image consists of slices of a contiguous sequence of pixel macroblocks (MB). In MPEG, a sequence is divided into groups of pictures (GOP) providing random access as well as error resiliency and including three types of pictures:

> 1) <u>Intra coded pictures, or I-pictures</u> use only information present in the picture itself, and provide potential random access points and error robustness inside the video sequence. They use only spatial transform coding without motion data. The first picture inside GOP must be an I-picture, since it allows decoding at the start of any GOP.
> 2) <u>Predicted pictures, or P-pictures</u> are coded with respect to some previous I-picture or P-picture (forward prediction). They use MC to improve compression efficiency and serve as prediction reference for B-pictures as well as future P-pictures.
> 3) <u>Bi-directional pictures</u> use both a past and future frames as a reference (bi-directional prediction). So, their best compression comes from using past and future frames as reference increasing computational cost and delay.

An efficient transmission format for 3D TV sends one color video view with a per-pixel depth map. The depth data helps rendering virtual views in which the objects of the monoscopic color video have been shifted to positions and captured by a virtual camera parallel to the real one. Depth map represented as a grayscale video can be better compressed than the corresponding color video while preserving good quality (Ozaktas & Onural, 2008). Stereo video is the most important special case of MVV with $N = 2$ views. Compression of conventional stereo video has been studied for a long time and the corresponding standards are available.

In this text, the problem of allocating bits to color and depth information when using H.264/AVC video codec is considered, because of the strong relationship between MCME and depth information in 3D TV. The quality of 3D videos stored as video plus depth map is affected by the number of bits used for coding color and depth when using baseline H.264.

Motion video plus depth (MVD) standards are still under examination by the JVT. The MPEG-C Part 3 and the MVC specifications do not handle well the MVD data representation and they may require extensions or combinations of other characteristics (Ozaktas & Onural, 2008).

### A.2.2 Video Quality Metrics

There has been a growing interest for better video quality metrics to automatically predict the perceptual improvement of video streams, video databases and numerous other applications. Motion-JPEG – where each frame is stored with the JPEG format - extends the JPEG standard to video.

Visual metrics are crucial in quality monitoring/assessment for broadcasting, compression, coding, EC and resiliency and they influence the entire design of video systems. There are two broad classes of quality measures: objective and subjective.

Objective video quality metrics rely on meaningful mathematical procedures without viewer panels and they are classified as follows: (1) Full-Reference (FR) metrics have full access to the original (reference) signal; (2) No-Reference (NR) metrics help when the reference signal is not completely accessible; and (3) Reduced Reference (RR) metrics use features from the original video and send them as side information to the receiver to help evaluating distorted video.

Subjective metrics try to imitate human vision, leading to improved methodologies since human beings are the ultimate video receivers. Nevertheless, a deficient human vision system (HVS) comprehension and difficulty in learning from physiological/psychological data degrade the performance of this class of metrics (Furht, 2008). Human vision is exceptional, but it seems to give unequal consideration to all data/cues, tending to center on some image parts. Visual awareness is central to the HVS and it is helps signaling saliencies amid objects in a scene (Kienzle, Schölkopf, Wichmann, & Franz, 2007). Progress achieved on subjective 3D quality evaluation have limitations and should be applied to specific features (for instance, compression artifacts) due to the required computational costs and processing time. The mean opinion score (MOS) i.e. has been regarded as the most reliable quality metric (Wang & Bovik, 2006), but it is cumbersome, and time consuming.

Spatial information (SI) features are related to the standard deviation of edge-enhanced images and it presupposes that compression will alter the edge statistics in the frames. Temporal information (TI) features are related to the standard deviation of the difference amid frames. Relationships between the reference and the distorted videos are derived from SI and TI features.

Many simple objective image metrics such as the PSNR have been used, but they fall short from the human perception of the differences involving frames. PSNR is also extremely sensitive to depth map errors. Perceptually-guided rendering focuses on fast photo-realistic algorithms to reduce the number of unnecessary computations. It is difficult to recognize 3D objects, because they can be represented in numerous ways such as polygonal meshes, parametric surfaces and discrete models.

Validation is an important step towards successful development of practical video quality metrics because it is indispensable to build a video database with subjective evaluation scores associated with each of the video sequences in the database. MCME methods require several evaluation metrics to help assessing and correcting dynamic scenes. Subjective and objective quality metrics are very correlated in 3D video. Hence, individual objective appraisals of 3D video can replace subjective procedures for most parameter alterations. These potential objective metrics are FR methods where the receiver needs the original 3D stream for quality assessment of the rendered video (Brunnström & al, 2009; Starck, Kilner, & Hilton, 2008).

Visual attention metrics encompass schemes to lessen the computational cost of the searching processes inherent to visual perception algorithms and they are modeled after a bottom-up process (BuP) based in image traits stimuli or a top-down task/comprehension (TDT) strategies. For a BuP, saliency levels are constrained by image characteristics/stimuli rather than visual information and, thus, creating a saliency map. The resulting sum of BuP responses can be nonlinear due to miscellaneous stimuli sources. The corresponding analysis has been done by subjective tests and modeled accurately for combinations of cues such as orientation, motion, luminance, color and contrast (Furht, 2008; Kienzle, Schölkopf, Wichmann, & Franz, 2007).

### A.3 the Role of MCME in an End-to-End 3D TV System

This section shows how MCME can improve all parts of an end-to-end 3D TV system, namely: image acquisition/creation, video coding/compression, broadcasting, reception, decoding/decompression, EC, and display/rendering.

### A.3.1 Acquisition and Content Generation
### A.3.1.1 Single Camera and 2D to 3D Conversion

The available techniques for 3D extraction from single-camera sequences are useful for conversion of the legacy mono-view video to the 3D TV. The compatibility with conventional 2D TV to ensure a gradual transition to 3D is referred to as backward-compatibility. 2D to 3D conversion suffers due to scarce information on depth and it relies on multiple cues from monoscopic video such as motion parallax and texture. Depth extraction methods must be content-adaptive and the most successful is 3D stereo video generation by the DIBR technique (Cheng, Li, & Chen, 2010). The translation of 2D material to 3D plays an important role in the implementation of 3D TV and it involves a backward-compatible system. The concept of depth maps takes advantage of the HVS ability to merge reduced disparity data that situated largely on boundaries where depth cues create an improved depth sensation over legacy 2D images.

Depth estimation based on displacement data may have uncertainties thanks to the simultaneous dislocation of objects and camera and to the accuracy of H.264-estimated MVs. ME in H.264/AVC and other standards rely on maximizing compression efficiency at the expense of the accuracy of object displacement estimates. Hence, not all MVs can be used to correctly estimate depth, but only those MVs that correspond to the object displacement (Pourazad, Nasiopoulos, & Ward, 2009).

In case of depth uncertainty, an object with larger motion than others is not unavoidably close to the camera, because its direction in relation to an object may have changed. If there is camera panning, then the predicted MVs corresponding to stationary regions are equal to the camera motion. These stationary regions are repeatedly classified by the H.264/AVC motion rating procedure such as the 'Skip Mode'. The transformed coefficients and the displacements are not transmitted once a block is omitted. Thus, the median of the MVs of the nearby blocks - identified as predicted MV - is used as the block MV. Besides panning, camera zooming can also cause depth ambiguity. If there the MVs are biased, then the estimated MVs are scaled accordingly. Zooming in/out may reverse depth or may cause eye weariness if not corrected during depth estimation. Nonlinear scaling models can be used for the reason that there is a nonlinear relation involving visual depth perception and the distance of an object. The suggested scaling factor reduces as the object distance increases – with superior depth sensitivity – and, hence, boosts the distinction amid depth values. The estimated depth map combined with the 2D video stream can facilitate the recovery of stereoscopic pairs via a DIBR algorithm in order to handle occlusion.

The low-cost single-camera technique solves the problems in mono-view video content and shows potential as a link between the available conventional video sequences and the 3D displays. The SfM scheme is considered the finest single-camera method for acquiring real life video. SfM is also adequate to human face and body capture. However, implementing synchronized multi-camera acquisition increases hardware complexity and makes the 3D TV system more expensive. Several algorithms have been proposed for creation of dynamic scene representations from the data acquired from the multi-view points. In this aspect, the virtual FVV generation from arbitrarily chosen viewing angles is an important step for 3D TV displays.

Techniques to retrieve 3D video such as monoscopic video, multi-view or stereoscopic vision result in ill-posed problems due to the conversion procedures. Although there is no ideal solution to this extraction problem, SfM is very popular and it tries to identify the 3D geometry involving camera and objects. Feature correspondences obtained from two (or more) video frames are required to establish geometrical relationships, to estimate camera calibration parameters or to infer scene structure (Schmid, Mohr, & Bauckhage, 2000). The converted 3D TV lacks the camera intrinsic structural parameters (Hartley & Zisserman, 2003). Structure and motion problems require a camera array with high precision and resolution thanks to the nonlinearity of the rendering process. Thus, formulations robust to motion variations should be favored.

### A.3.1.2 Multi-Camera Systems

The majority of multi-camera systems for 3D TV and 3D reconstruction are intended for telepresence, telemedicine and teleconferencing. This type of capture system requires software and/or hardware synchronization (Yang, Everett, Buehler, & Mcmillan, 2002) as well as calibration. The last step involves a more complex correspondence problem, since there are multiple parameters to be found (a set of parameters for each camera). A dense camera mesh captures better the light field; however high-quality reconstruction filters could be used, provided the light field is undersampled (Stewart, Yu, Gortler, & Mcmillan, 2003).

### A.3.1.3 Model-Based Systems

Various views from a camera array can be used to interpolate 3D image in order to render it properly in the synthesis stage (Carranza, Theobalt, Magnor, & Seidel, 2003; Fehn, et al., 2002; Zitnick, Kang, Uyttendaele, Winder, & Szeliski, 2004). Throughout rendering, the MVV can be combined with a mathematical and/or computational model to build better quality surfaces (Carranza, Theobalt, Magnor, & Seidel, 2003). Real-time capture of high definition video relying on models is complex due to needs such as extra storage and transmission bandwidth.

The multiple-frame structure from motion (MFSfM) involves more than three frames obtained with a monocular single-camera which is circles the desired subjects. The MFSfM requires the existence of causality and temporal continuity constraints, so that there is a small disparity between frames. Temporal redundancy is of limited use, without excellent feature tracking and inter-frame correspondence. Although linear algorithms give straightforward solutions, the SfM problem is intrinsically nonlinear and dependent on linearization assumptions (Kanatani, 1990). In a multi-frame context, the main algorithms rely on the fact that image measurements can be expressed as a product of two matrices, representing the motion and the structure. Nonlinear algorithms also depend on iterative minimization of a cost function on both structure and motion parameters Nonlinear methods rely on an iterative cost function minimization of a cost function on both structure and motion parameters. Local minima, nonguaranteed convergence and dependency on good parameters initialization are some limitations of these algorithms.

Linear and nonlinear cost functions can be proposed to complete absent entries, handle occlusions and unreliable features (Jebara & Pentland, 1999). Fusion methods estimate structure by combining intermediate reconstructions resulting from smaller subsets of the sequence. Another strategy for SfM considers state estimation for dynamical systems with new constraints and reducing the solution space (Oliensis, 2000).

The use of 3D images and model databases is growing, but fall short to match users' expectations. Efficient data recovery from collections of images and videos involves offline and online processes for 3D model indexing. Offline processes use performance descriptors that are invariant to particular transformations. In online processes, the end-user queries the search engine using similarity metrics to compare 3D objects, for a given model description (Stoykova, et al., 2007). Possible descriptors are:

• Statistical approaches focus on the probability distributions of 3D local and global model descriptions.
• Structural approaches look for high-level structure, such as a graph, from a 3D mesh.
• To handle 3D shapes in other domains with the help of transformations.
• To represent a 3D model by a set of 2D-decribed characteristic views.

Many 3D applications are limited to human face or full body processing. They involve face and facial feature detection; capturing of 3D structure of the face; analysis of global face motion/mimic; 3D modeling; kinematics and motion analysis; and motion recognition. Algorithms can be more efficient with a priori knowledge about 3D structure and motion of human faces and bodies which is extremely necessary for the development of 3D visualization technology. In 3D visualization and 3D display systems, robust detection and tracking of the observer's eyes and the observer's view point is necessary to render the correct view according to the observer position.

### A.3.2.1  3 D Compression and Coding

When it comes to high-quality rendering of 3D video and FVV, MPEG has been revised to include more possibilities (Smolic & McCutchen, 2004), because of the existing challenges (Bourges-Sevenier & Jang, 2004). MPEG-C Part 3 handles video plus depth data (ISO/IEC JTC1/SC29/WG11, 2007). Transmission and 3D TV with video plus depth can be also be found in (ITU-T Recommentation H.264 & ISO/IEC 14496-10 AVC, 2005). Additionally, H.264/AVC express depth through its auxiliary picture syntax Martinian, Behrens, Xin, Vetro, & Sun, 2006). Figure A.4 depicts a general 3D TV content producer. The latest H.264/MPEG-4 AVC standard incorporate features from the MPEG-2 codec with a mixture of temporal and spatial estimation (Micallef, Debono, & Farrugia, Sep. 2010) characteristics, as follows: temporal prediction with variable block sizes, multiple reference frames, intra prediction, adaptive entropy coding, and filters to reduce the blocking effect. Moreover, it uses more efficiently the spatio-temporal correlation between adjacent MBs with the SKIP mode (predictive P-slices) and the DIRECT mode (bi-predictive B-slices) (Schwarz, Marpe, & Wiegand, 2006). Compression and prediction of 3D OFF can reuse past coded MVs. If there is some motion information that can been skipped without compromising image retrieval at the decoder side, then the bitrate savings may be higher.

The MPEG-7 standard provides mechanisms to represent audiovisual data in multimedia systems (Manjunath, Salembier, & Sikora, 2002; Sikora, 2001): descriptors (D), description schemes (DS), description definition language (DDL) and additional features. Descriptors symbolize parts of the syntax and the semantics of each attribute representation (Manjunath, Salembier, & Sikora, 2002). DS, schemes detail structure, semantics and the relationships among components (D and DS). DDL allows the DS and maybe descriptors creation/modifications; system tools to support multiple descriptions; synchronization; transmission methods; file formats; etc...

Multimedia descriptions encompass still pictures, video, graphics, audio, 3D models and their relationship. It is desirable that a 3D descriptor also supports any 2D descriptors - such as contour shape, color or texture –used to render real-world objects; c.g. (ITU-T  H.264 & ISO/IEC 14496-10 AVC., 2005).

Unmistakably, excellent MVV coding is vital to 3D TV. The simplest way to solve the multi-view coding problem is to code video streams independently in time and transmit each of the views, entitling the use of current video coding standards and codecs. Conversely, the coding efficiency is for the most part lower than with other multi-view coding methods.

MVV compression has habitually focused on static light fields e.g., (Magnor, Ramanathan, & Girod, 2003). There is a need to research in compression and transmission of MVV in real-time (Yang, Everett, Buehler, & Mcmillan, 2002). Most systems compress the MVV off-line and focus on providing interactive decoding and display (Javidi & Okano, 2002). MC in time is called temporal encoding, and disparity prediction between cameras is called spatial encoding (Tanimoto & Fuji, 2003). Combining temporal and spatial encoding also leads to good results (Zitnick, Kang, Uyttendaele, Winder, & Szeliski, 2004). An extra approach to MVV compression, endorsed by the ATTEST project (Fehn, et al., 2002), is to condense the data to a single view with per-pixel depth map, compressing these data in real-time and broadcasting them as an MPEG enhancement layer. At the receiver, stereo or multi-view images are created using image-based rendering, but high-quality output is impaired by occlusions or high disparities in the scene (Chen & Williams, 1993). Furthermore, one view is not enough to capture view-dependent effects, such as reflections and highlights.

In high-quality 3D TV distribution, all views are sent to many users at once. For compression and transmission of dynamic MVV data, either the data from multiple cameras is compressed using spatial or spatio-temporal coding, or each video stream is compressed by temporal coding. The first option gives higher compression, due to the high coherence between the views, but it requires the compression of multiple video streams by a centralized computer, which is not scalable, since inserting extra views surpasses the coder bandwidth.

An extra benefit of using existing 2D coding standards is that the codecs are well established and broadly accessible. Tomorrow's 3D TV could include one or many decoders, depending on the amount of views available for an object or scene. A system should be complaint with several 3D TV compression algorithms, allowing for multiple views vin all its stages.

In order to design 3D video (3DV) and FVV, the entire 3D processing chain needs to be considered (Ozaktas & Onural, 2008), because of the existing dependence among its elements. For instance, an interactive display requiring random access to 3D data will shape the performance of a coding scheme based on data prediction. There are several different 3D scene representations, which result in numerous data types.

Since both stereo pair images are very similar, one image can help predicting the other. The second image can be predicted from the one already coded, similarly to the way temporally related images can be motion-compensated in video compression. The samples of both images relate to each other through the 3D scene geometry and camera properties, including positions and internal camera properties such as the focal length. The displacement or disparity of each sample in one image with respect to the other is equivalent to a dense motion field in between consecutive images of a video sequence. Thus, MEMC helps disparity estimation and disparity compensation in image prediction followed by coding of the prediction error or residual.

Some differences between motion compensation and disparity compensation exist due to the dissimilarity between the statistics of disparities and MVs. Disparities are biased and larger than MVs. Then, small disparity means large depth of the corresponding point in 3D. 3D points close to the camera may have a very large disparity value which calls for modifications of entropy coding of the disparity vectors. Additionally, temporally adjacent video images tend to be more alike than views of a stereo pair at realistic frame rates. In general, disocclusion effects are more apparent in a stereo pair than in two video frames next in time. Moreover, some differences in a stereo pair result from erroneous white and color balance, but also due to scene lighting and surface reflectance effects.

The combination of inter-view and temporal prediction is the central idea behind efficient stereo video compression ruled by the standard defined in ITU-T Rec. H.262 | ISO/IEC 13818-2 MPEG-2 Video, the Multi-view Profile (Garbas, Fecker, Troger, & Kaup, 2006).Typically, the increase in compression performance compared to independent coding using both video streams is not noteworthy because temporal prediction efficiency is already admirable. In general, if time prediction is excellent further inter-view prediction does not boost coding efficiency a lot. Images adjacent in time tend to have more similarities than spatial adjacency.

I pictures are coded without reference to other temporally adjacent images in the dynamic sequence; a great gain can be accomplished by inter-view prediction. Inter-view prediction coding increases compression efficiency considerably compared to coding this picture as I picture.

Compression of stereo video may include optimum joint bit allocation for both channels, in detriment of backward compatibility to plan better inter-view prediction structures. Algorithms for modern video codecs must take into account new and current standards such as MPEG-4 Visual and H.264/AVC along with HVS understanding and stereo perception.

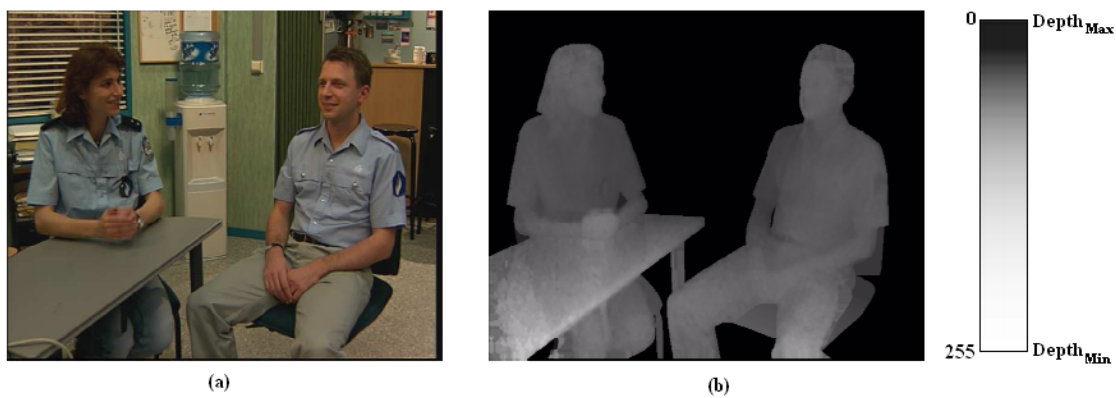

*Figure A.5   – Video Plus Depth Format for the Interview Sequence consisting of (a) a 2D color frames with  (b) a depth frame. Depth is code with 8 bits.*

Compression of video plus depth data is as an option to typical stereo video, where the stereo pair is generated by view interpolation. This format increases compression performance and depth can be considered as a signal with just monochromatic and luminance information. Fig. A.4    shows a 2D color image with its corresponding depth map. The depth ranges from $Depth_{Min}$ (close to the viewer) and $Depth_{Max}$ (far away from the viewer) as a gray scale image with linear distribution from 255 to 0 in that order. These gray values can be inputted to the luminance channel and the chrominance can be made constant. Hence, the resultant signal can be handled by any video codec. This format has been investigated by the European ATTEST project with several state-of-the-art video codecs (MPEG-2, MPEG-4, H.264/AVC). Depth can be coded at high quality with merely 10–20% of the bit rate because the depth data statistics is smoother and less structured than color data. Indeed, even strong depth-image coding problems such as blocking artifacts do not correspond to perceptible distortions in the rendered views.

Both 3DV and FVV systems use multiple views of the same scene that have to be sent to the user. A simple solution would be to code all signals independently using a state-of-the-art video codec like H.264/AVC . MVV contains lots of inter-view statistical dependencies because all cameras capture a scene from different viewpoints. These can be exploited for combined temporal/inter-view prediction images are not only predicted from temporally neighboring images but also from corresponding images in adjoining views. Investigations in MPEG have shown that MVC algorithms surpass independent coding with gains of more than 2 dB for the same bit rate.

Any video coding standard must have a high compression efficiency. For MVC, the performance should be higher than when independent compression is employed. Small computational costs, error robustness, support for different pixel/color resolutions and low delay are also important requirements. A standard mechanism called Profiles may help satisfying these application prerequisites.

MVC requires temporal random access which can be obtained with the help of I frames. Scalability is a sought-after attribute for most video coding standards because a decoder can access part of a bitstream to output a low-quality video which reduces temporal or spatial resolution, or a reduced video quality. For MVC, extra view scalability is necessary by using a bit stream fraction in order to yield a limited number out of the N views. Also backward compatibility is vital for MVC. Thus, one bitstream corresponding to one view that is extracted from the MVC bitstream shall be compliant to H.264/AVC. It should be feasible to fine-tune coding to equalize quality over all views. Parallel processing allows for a great coder realization and resource management. Both extrinsic and intrinsic camera parameters should be sent with the bitstream to provide intermediate view interpolation at the decoder.

The key MVC aspiration is to boost compression in contrast to coding all video data separately. It is a common practice to code all views with H.264/AVC independently and to use the result for objective and subjective performance assessments. Coding can be done using common settings and parameters. As a rule, illumination and color inconsistencies can be reduced by taking into account light variations over multi-view images thanks to the illumination circumstances. The fundamental idea is to modify MC on macroblock level, assuming locally constant illumination and color. Prior to subtracting the sample values of the block to be coded and the reference block, their means are compensated. Illumination correction has been adopted by the Joint MVV Model (JMVM) (Vetro, Su, Kimata, & Smolic, 2006) as an option. Other strategies are working on macroblock level at the coding stage and pre-processing prior to coding. Since algorithms for light compensation are well-known, then the corrected data can be assigned to a typical coder without the need to project a new coder/decoder and bitstream syntax.

Disparity coding may improve inter-view prediction. An alternative to improve disparity estimation is to treat it like motion, although its statistical properties can differ from MVs and to take geometric properties and constraints into account. Specific coding modes for MVC such as the inter-view direct mode are also under research (Guo, Lu, Wu, & Gao, 2006).

High-level syntax is required to indicate the properties of a MVC bitstream to a decoder. Extra high-level syntax is under improvement to permit resourceful random access, buffer supervision, and parallel processing. Additional paths in MVC analysis like combining scalability and MVC are under investigation, but in general scalability lessens compression efficiency.

Suitable video coding schemes for compression and transmission that are based on the 3D video formats are discussed in this part of the text. Since pictures from a stereo video sequence are very similar, then they are well suited for compression with one predicting the other.

An alternative to MVV compression is to diminish the data to a single view with a per-pixel depth map in order to send them out using the MPEG enhancement layer. Stereo or multi-view frames rendered at the receiver may be impaired by view-dependent effects, occlusions and/or high-valued disparities. The 3D ATTEST chain can set apart distortions in video coding, depth quantization, and geometry. The MPEG committee has been examining numerous contributions to the video plus depth issue and techniques were proposed to allow for estimation over time and across space.

Constant translational motion in video corresponds to a planar Fourier spectrum, i.e., non-zero Fourier-transform amplitudes are located on a plane passing through the origin of the frequency space and orthogonal to the movement direction. This allows for the use of a speed voting strategy or filtering. When it comes to the discrete Fourier transform (DFT) similar behavior occurs.

In the case of a stereo system, a multi-view profile (MVP) has been defined in MPEG-2 standard, which allows the transmission of two video signals for stereoscopic TV applications. One of the main features of the MVP is the use of scalable coding tools to guarantee the backward compatibility with the MPEG-2 Main Profil. The MVP relies on a multilayer representation such that one view is designed as the base layer and the other view is assigned as the enhancement layer. Also, the MVP conveys the camera parameters (i.e. geometry information, focal length, etc.) in the bitstream. The base layer is encoded in conformance with the Main Profil, while the enhancement layer is encoded with the scalable coding tools. Temporal prediction only is used on the based layer, while temporal prediction and inter-view prediction are simultaneously performed on the enhancement layer. As a consequence, backward compatibility with legacy 2D decoders is achieved, since the the base layer represents a conventional 2D video sequence.

Stereo video information has to be organized, so that the decoder can distinguish the left and right view inside the bitstream. The stereo video supplemental enhancement information (SEI) message defined in H.264/MPEG-4 AVC fidelity range extensions (FRExt) helps implementing this mechanism.

The MPEG-C Part 3 (aka ISO/IEC 23002-3) standardizes video-plus-depth coding based on the coding of 3D content inside a conventional MPEG-2 transport stream, which includes texture, depth and some auxiliary data. This solution provides interoperability of the content, display technology independence, capture technology independence, backward compatibility, compression performance and the ability of the user to control the global depth range without increasing the bandwidth too much.

An additional coding tool for video-plus-depth data is the multiple auxiliary components (MAC) defined in version 2 of MPEG-4. The MAC is not only used to describe transparency of video objects, but can also describe shape, depth or texture. Therefore, the depth plus video can use the auxiliary components. MAC coding uses motion compensation and DCT like it is done with the MVs of the 2D video. This coding scheme represents the disparity vector field, the luminance and the chrominance data by 3 components.

Some CG research resulted in the animation framework extension (AFX) of MPEG-4 which keeps backward compatibility. Three tools of interest in the scope of 3D video are depth image-based rendering, point rendering and view-dependent multi-texturing.

The general case for two or more views appears in the draft MVC specification that provides new ways to improve coding efficiency, to reduce decoding complexity and to diminished memory consumption for various applications. The compressed multi-view bitstream include a base layer stream that could be easily extracted and used for compatibility with legacy 2D devices.

For each camera, view dependencies are transmitted using the sequence parameter sets (SPS) MVC

syntax which improves MVV content compression exploiting redundancies among the inter-pictures of one camera and the inter-view pictures of other cameras, and leads to an additional coding gain compared to the H.264/MPEG-4 AVC simulcast.

### A.3.2 Decoding/Decompression

The receiver generates all potential views (the entire light field) to the user all the time. The display controller asks for one or more virtual views by stating the virtual cameras parameters. The decoders receive compressed video, and store the decoded frame in a buffer (see Figure 4). There is a virtual video buffer (VVB) for each end user with data from all received source frames. A user watches a rendered scene generated after processing pixels from several frames in the VVB. Due to bandwidth and processing restrictions, each end user cannot obtain the entire source frames from all the decoders. This also decreases the system scalability.

When views are decoded from an MVC bitstream, some views may not be displayed, but are needed for inter-view compensation and decoding of the target views. The single loop decoding (SLD) method from the MVC requires partial decoding of reference views and it relies on the inter-view SKIP mode. Then, decoding temporal non-key pictures in reference views equals to not fully decode MBs, but to storing motion information to decode the corresponding MBs in the target views.

### A.3.4 Error Concealment (EC)

The consequences of spatial impairment and the use of multiple reference picture estimation in both spatial and temporal dimensions are still being examined (Farrugia & Debono, 2010). EC mechanisms exploiting the spatial and temporal coherences among views are needed in order to: (i) attain the best possible efficiency in coding/compression, (ii) correct light and color, (iii) lessen residual errors caused by spatial estimates (Coelho, Estrela, & de Assis, 2009), and (iv) account for cameras uncertainties (Stoykova, et al., 2007). H.264/AVC finds displacements by favoring compression efficiency over displacement precision. So, two blocks with identical MVs may not belong to the same object or part of the object under analysis which leads to errors in disparity and MV estimation (Micallef, Debono, & Farrugia, Sep. 2010).

Joint-source and channel-coding strategies developed for video transmission over channels can alleviate packet losses in streaming video due to limited bandwidth. Nevertheless, EC methods are needed at the receiver to reduce the observed damages caused by these losses. In general, if a packet is lost, then the entire video frame is affected and this is a very notorious drawback in video streaming. The onward MVs of the last received frame can be estimated via methods that exploit knowledge from precedent frames. That is, EC uses MVs to improve the quality of the damaged frame by inferring the missing blocks and rendering an approximation of the real frame.

Conversational applications have benefited from EC algorithms a lot. However, video streaming applications have been less successful when it comes to error remediation up to now. If a packet vanishes, probably a complete video frame is gone, so that the majority of the EC algorithms fail (Micallef, Debono, & Farrugia, 2010). The most relevant problems related to EC are the following: i) Lost MVs cannot be estimated from neighboring blocks, if the corresponding MVs are not available; ii) Without nearby MVs adjoining MBs do not help to re-estimate the MVs via block matching with respect to a reference frame; iii) one cannot use the nearby MBs to test the side boundary match of a candidate substitute MB; iv) partial decoding of DCT coefficients is not offered; v) syntax-based repairs are not useful as a result of the bursty nature of the errors; vi) even spatial EC is not feasible because no neighboring MBs is available. So, video streaming applications need an EC algorithm that can handle loss of entire frames.

EC algorithms tailored for video streaming should estimate whole missing frames of a sequence and exploit the multiple reference frame buffer provided by H.264 to reduce the recovered video error. Specifically, the algorithm is based on the estimation of the absent frame MV field, the previous received frames, and on the previous frame projection on the missing one founded on this estimated data.

The difference among viewpoints inside a MVV may provoke illumination changes between sequences. Inter-view illumination changes between the pictures can be balanced by means of illumination change-adaptive MC (ICA MC). The MVC extension uses the multiple reference frames feature of H.264/MPEG-4 AVC.

Inter-view correlations combine motion/disparity compensated estimation, where a predictive frame is created from temporally adjoining frames. As an extension of the traditional temporal SKIP mode, inter-view SKIP mode comes from the idea of exploiting the relationship between MVs between neighboring views. So, the motion data of the current MB is derived from the corresponding MB in the image at the same temporal index of the neighboring view.

## A.3.5 The role of MCME in 3D image rendering
### A.3.5.1 **Multi-view Scene Reconstruction**

Multi-camera systems capture dynamic scenes from numerous viewpoints. Multi-view scene reconstruction can be performed by: 1) minimization of a cost function related to the 3D volume; 2) iterative evolution of a surface, where the rendering quality is associated to cost-function optimization; 3) mixing individual views together; 4) reconstruction from feature points (surface fitting); and 5) reconstruction from silhouettes.

3D TV implies depth impression of the observed scenery and FVV allows interactive choice of viewpoint and direction in a certain range as well-known from CG. They can be combined in a single system, provided they are based on a suitable 3D scene representation. If a stereo pair can be rendered, then 3DV is feasible. If a camera view not normally accessible (virtual view) related to a random viewpoint can be rendered, then FVV is viable.

The definition of a 3D representation format is essential to improve rendering algorithms, navigation range, quality, interactivity, compression and broadcast for 3D TV and FVV. Image-based representations call for dense camera arrays in order to render properly virtual views.

Photo-realistic 3D scene and object modeling is often difficult and time consuming. An automatic 3D object and scene reconstruction implies camera geometry estimation, depth structures, and 3D shapes. All these methods introduce geometric rendering errors. Another challenge in 3D scene representation is image-based modeling which requires intermediate virtual views obtained from the interpolation of available camera views and/or dense sampling with several camera views.

3D is perceived because the information from each eye about a slightly different viewpoint of an object is combined by the brain and it creates the sensation of depth. This property is used to design 3D displays. Two cameras capture scenes from distinct viewpoints as if each of them was an eye. If a display allows each eye to see only its related view, then a 3D impression is created and it can be captured in stereo or converted from existing 2D video allowing for 2D to 3D conversion.

The FVV functionality makes possible to implement depth-based stereo rendering. A video signal and a depth map are transmitted to the user, helping rendering a stereo pair. Depth and disparity estimation can only be solved up to a residual error probability because data is in general incomplete. Estimation errors affect the quality of the generated views. One method is to use a stereo camera for acquiring and estimating depth/disparity from correspondences or from motion as well as other single view video properties.

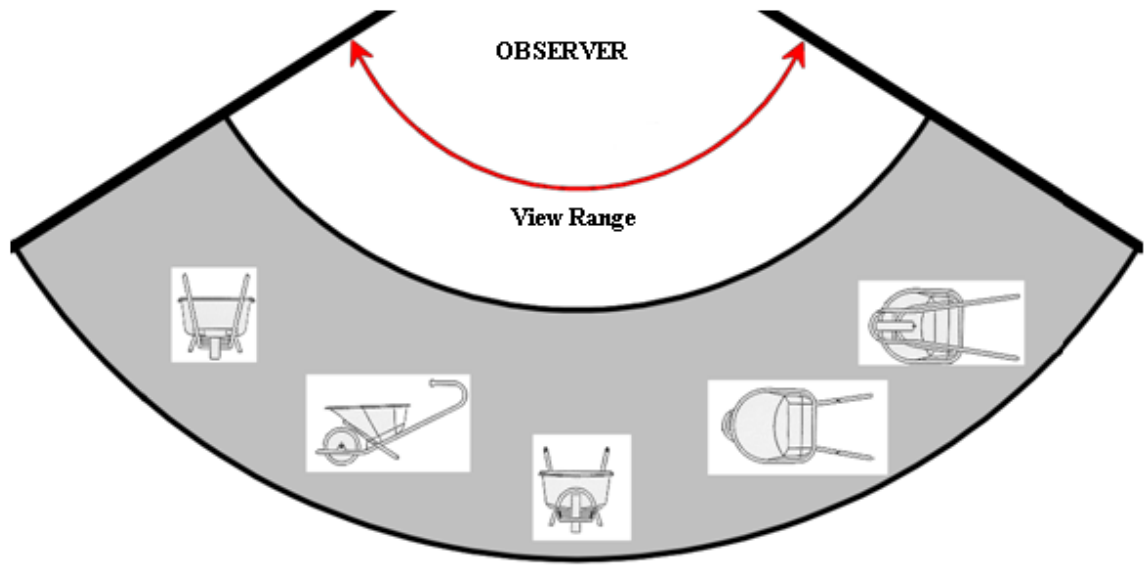

*Figure A.6 - FVV with interactive selection of virtual viewpoint and viewing directions. An object in shown with 5 radically different views (N=5).*

FVV offers the same 3D functionality as CG, that is, the user can select a viewpoint and viewing direction within a visual scene, meaning interactive free navigation. The figure above illustrates the complexity associated to these features. The red arc shows the amplitude of the observer's perception in a plane parallel to the ground, but the viewer can actually change his/her viewpoint in a 3D framework. A perfect viewing experience would require lots of scene interpolations and would slow down rendering in order to allow a natural 3D impression of the object whose orthographic views are shown above.

### A.3.5.2 Virtual/Augmented Reality

Because of convergence, future TV sets will permit navigation through a scene and will be able of handling more than movies and games. There are also several situations where either the application or the extent of the terrain under scrutiny demand distance measurements. Bearing these issues in mind, range imaging is an option, since each range image pixel has a property corresponding to a sensor that can be given in physical units. Distance and other properties can be tracked and mapped in real time which will improve considerably environmental and disaster remediation (Fernandes, do Carmo, Estrela, & Assis, 2009). The use of this type of equipment combined with photogrammetry can be valuable in remote sensing and surveillance (Gokturk, Yalcin, & Bamji, 2005) (Vlasic, et al., 2009).

It is easier to measure distance/depth with a time-of-flight (TOF) sensor than with stereo and triangulation. A small amount of processing power is used because the software required by a smart sensor is simple. Hence, computational effort can be put on more demanding tasks such as classification and target location. As TOF images can be obtained in real time, tracking human beings and their interaction with games/simulators is simplified. In automation, mobile robots can map their surroundings in order to evade obstacles or track a person or vehicle. TOF techniques can also replace radar or LIDAR with a smaller computational load.

Motion capture technologies help rendering augmented-reality views because the output of a 3D camera can be combined with live creatures in CG environments (Sorbier, Takaya, Uematsu, Daribo, & Saito, 2010). Because TOF cameras capture a whole 3D scene with dedicated image sensors, the hardware complexity decreases (Gokturk, Yalcin, & Bamji, 2005).

### A.4 Solutions and Recommendations

High spatial resolution as well as efficient coding of depth and texture are technological challenges in 3D TV. Coding/compression performance advances can result from the strong correlation between the color video and the depth map sequences. The amount of information used to describe textured video motion with depth map and a common MV field need to be compacted in order to make the overall system behave as close as possible of real time. The bitrate control scheme should depend on the content of each sequence instead of having a fixed percentage range.

Handling 3D scenes poses a severe computational burden. When color is added to 3D video, the scenario becomes even more complicated. Hence, new formats to describe color objects and more efficient metrics to evaluate color distortion and geometry are needed.

Depth pre-processing along with depth-aided in-painting can diminish the occlusion/disocclusion effects.

It is indispensable to trim down temporal and inter-view redundancies of frames with joint motion/disparity field estimation in MVV while keeping motion and compression artifacts to a minimum. Segmentation and coding based on ratio distortion (RD) can replace the block-based motion/disparity estimation stage in the MVC extension, thus improving RD efficiency (Vetro, Su, Kimata, & Smolic, 2006). The coding performance vs. bitrate impasse for high-quality rendering needs further research (He, Yu, Yang, & Jiang, 2008). Furthermore, the stereo-motion consistency constraint means that all displacement vector fields are greatly correlated. Hence, they can benefit from interfield compression (Guo, Lu, Wu, & Gao, 2006). Several multi-view-based video compression schemes exploiting the intrinsic correlation in MVV need to be considered.

Since 3D video involves several views they can be coded independently which increases bandwidth compared to established coding techniques. However, prediction across views can improve compression performance. Bitrate requirements for 3D services can be considerably reduced if the HVS is taken into account. For instance, different views can be coded differently without spoiling the 3D sensation in way that resembles in the HVS asymmetric view. Hence, understanding the asymmetric coding will perk up the quality of asymmetrically coded video.

Better distortion models to describe the quality of the rendered views will also improve 3D video codecs. Since 3D perception quality is very important, ignoring the HVS poses grave problems.

All stages of a 3D end-to-end system can also benefit from research on domain transformations.

**FUTURE RESEARCH DIRECTIONS**

It seems quite impossible to acquire and to render high quality 3D TV content without combining computer vision and computer graphics. Hence, it is probable that the frontier between these areas of knowledge disappears and it will give birth to a new one encompassing convergence issues.

It seems that convergence is a much broader and more complicated issue than simply combining technologies. There is a need for "knowledge convergence" as well. Just to site an example, we have a tendency to think about images in the visual part of the frequency spectrum due to the fact that we as human beings "see" the world using our eyes only. However, the internet of the future will carry multispectral and hyperspectral images. Moreover, the existence of tactual and olfactory data should not be forgotten.

Standards help developing better video systems from capture to display, although it is not possible to establish them and believe they can be always "extended" to adapt changes. Other standards must be studied/proposed in conjunction with the existent ones, since there is always the risk that too many amendments can compromise other technological solutions that may become more viable in the long term.

Interactivity needs to be handled not only from the event-driven point of view, but depending on the application, it has to take into consideration the end user's psychological responses to "change". Is the offline-to-online digital shift going to occur more often? Or maybe the right question would be "Is the offline-to-online digital shift going to happen?" Our point is, shall the end user be the only entity responsible for the content s(he) is exposed? We think that at some point in time real life will have to be encouraged.

Next-generation video infrastructure will rely on cloud computing. Ubiquitous access and the necessity of high quality video pose network restrictions such as computing ability, storage and optimal bandwidth.

## CONCLUSIONS

It's clear from the literary reviews in this chapter that 3D technology demands more improvement. Video systems for 3D are complex and have room for lots of progress. Besides the obvious issues related to transmission and reception of such content, both acquisition and reception suffer with the lack of fast and efficient algorithms, adequate structures for capturing and displaying images as well as more intelligent computation units.

Independently from the technology used, it seems that all stages do need better ways of summarizing information without sacrificing quality. This text tackles this problem looking from the visualization and content generation perspectives.

Motion data play an important role in all stages of an end-to-end 3D TV system. The tight relationship between motion and depth impacts these stages because 3D information can be recovered from these cues, and they help optimizing the performance of all algorithms and standards from different parts of 3D system.

Current 3D video coding standards are mostly extensions of present 2D standards. There is still room for lots of research on 3D video data representations since more standards for interactive 3D video services need to be established. Global motion field estimation, coding and compression can exploit the correlation between sequences and carry out a better bit allocation strategy. View quality can benefit from knowledge about the compression impact on the rendered scene and the influence of depth in video compression.

Depth pre-processing helps lessen geometric distortion for small baselines. On the other hand, large baselines amplify geometric distortions. a post-processing approach to handle large baselines and disoccluded regions has been suggested by (Daribo, Miled, & Pesquet-Popescu, 2010). Image inpainting (alias image completion (Furht, 2008)) fills in pixels in image regions where they are absent or distorted with the help of the neighboring data and relates to error concealment. Since the depth map is a 2D representation of a 3D scene, it is relevant to appraise the depth compression-induced artifacts on the 3D warping process. Preserving discontinuities is essential in depth data compression, because lots of 3D image effects demand robustness to abrupt changes in the depth map. The impact of depth map compression on rendering is analyzed by the MPEG 3DAV AHG activities (Schuur, Fehn, Kauff, & Smolic, 2002) with an MPEG-4 compression scheme. They applied decoding combined with a median filter to limit the coding-induced artifacts typical of view synthesis.

Depth map coding in MVV sequences decomposed with motion-compensated wavelet decomposition and multiple auxiliary components from MPEG-4 MAC are not closed problems and need more developments.

The joint MV field is part of the MPEG-2 texture stream for backward-compatibility purposes. Transmitting the correlation among MVs with both texture and depth streams can enforce error resiliency of a video-plus-depth sequence, because the lost data in one stream can be compensated by the information from the other.

Dense disparity estimation followed by block-based segmentation and coding in MVC generally produces a smooth disparity field which increases precision. Enforcing the smoothness constraint to these estimates improves SKIP prediction, and consequently, coding performance. It is indispensable to enforce displacement vectors consistency in a stereo/multiview sequence for joint motion/disparity estimation. The resulting model can be solved using convex optimization. An H.264-based codec can split each dense motion/disparity subfield into variable block sizes, while keeping the RD cost minimum.

Temporal/inter-view redundancies of key and non-key frames need to be reduced and disparity estimation and joint motion/disparity estimation are very useful. A possibility is to estimate dense motion/disparity fields followed by RD-driven segmentation and coding. This could replace the block-based motion/disparity estimation stage in MVC extension, improving the overall RD performance. Research on coding improvement in terms of bitrate and the quality of the synthesized image will result in better depth map quantization procedures.

DIBR helps synthesizing new virtual views from the video-plus-depth data. DIBR main bottleneck is the occlusion/disocclusion problem. This calls for efficient smoothing of the sharp depth changes near object edges. Geometric distortions and the computation time can be reduced by means of uniform filtering of the depth video. Large disocclusions can be handled by post-processing using depth data combined with inpainting techniques.

High-dynamic range imaging is a promising research area to help reducing the computer load related to 3D TV.

## REFERENCES

Belfiore, S., Grangetto, ,. M., Magli, E., & Olmo, G. (2003). An error concealment algorithm for streaming video. IEEE Intl. Conf. on Image Proc. 2003 (ICIP 2003 ), 22, pp. 649-652.

Bourges-Sevenier, M., & Jang, E. (2004). An introduction to the MPEG-4 animation framework extension. *IEEE Trans. on Circuits and Systems for Video Technology , 14*, pp. 928–936.

Brunnström, K., & al, e. (2009). VQEG validation and ITU standardization of objective perceptual video quality metrics. *IEEE Signal Proc. Mag. , 96*.

Carranza, J., Theobalt, C., Magnor, M., & Seidel, H. (2003). Free-viewpoint video of human actors. *ACM Trans. on Graphics , 22* (3), pp. 569–577.

Chen, S., & Williams, L. (1993). View interpolation for image synthesis. *SIGGRAPH 93 Proceedings*, (pp. 279–288).

Cheng, C., Li, C., & Chen, L. (2010). A 2D to 3D conversion scheme based on depth cues analysis for MPEG videos. *IEEE Trans. on Cons. Electronics , 56* (3), pp. 1739-1745.

Coelho, A., Estrela, V. V., & de Assis, J. (2009). Error concealment by means of clustered blockwise PCA. *Picture Coding Symposium.* Chicago, IL, USA: IEEE.

Coelho, A. M. Estrela, V. V. (2012a). Data-Driven Motion Estimation with Spatial Adaptation, *Intl. J. of Image Proc. (IJIP)*, Vol. 6, Issue 1, pp. 53-67. Retrieved in May 7, 2012 from http://www.cscjournals.org/csc/manuscript/Journals/IJIP/volume6/Issue1/IJIP-513.pdf

Coelho, A. M. Estrela, V.V. (2012b). EM-Based Mixture Models Applied to Video Event Detection, *In Principal Component Analysis - Engineering Applications*, Intech, pp. 102-124. Retrieved in May 7, 2012 from http://www.intechopen.com/books/principal-component-analysis-engineering-applications/em-based-mixture-models-applied-to-video-event-detection

Daribo,. I., Miled, W., & Pesquet-Popescu, B. (2010). Joint depth-motion dense estimation for multiview video coding. *JVCI Special Issue on Multi-Camera Imaging, Coding and Innovative Display: Techniques and Systems* .

Deng, X., Jiang, X., Liu, Q., & Wang, W. (2008). Automatic depth map estimation of monocular indoor environments. *2008 Intl Conf. on Multimedia and Inf. Tech. (MMIT 2008)*, (pp. 646-649).

Estrela, V. V., & Galatsanos, N. (2000). Spatially-adaptive regularized pel-recursive motion estimation based on the EM algorithm. *SPIE/IEEE Proc. of the Electr. Imaging 2000 (EI00)*, (pp. 372-383). San Diego, CA, USA.

Farrugia, R., & Debono, C. (2010). Resilient digital video transmission over wireless channels using pixel-level artefact detection mechanisms. In F. De Rango (Ed.), *Digital Video* (pp. 71-96). Intech.

Fehn, C. (2006). *Depth-image-based rendering (dibr), compression, and transmission for a flexible approach on 3DTV.* (B. M. Technical University, Ed.) Berlin, German.

Fehn, C., Kauff, P., De Beeck, M., Ernst, F., Ijssel- Steijn, W., Pollefeys, M., et al. (2002). An evolutionary and optimised approach on 3D-TV. *Proc. of Intl. Broadcast Conf.*, (pp. 357–365).

Fehn, C., Kauff, P., De Beeck, M., Ernst, F., Ijssel-Steijn, W., Javidi, B., et al. (2002). *Three-dimensional television, video, and display technologies.* Springer-Verlag.

Fernandes, S. R., do Carmo, F., Estrela, V.V., & Assis, J. (2009). Using the SIFT (Scale Invariant Feature Transform) to determine pairs of image points for using in the SITH (3D Hybrid Imaging System). *Proc. of the XII Workshop on Comp. Modeling (XII EMC).* Volta Redonda, RJ, Brazil.

Flierl, M., Mavlankar, A., & Girod, B. (2006). Motion and disparity compensated coding for video camera arrays. *IEEE Proc. of Pict. Cod. Symp. (PCS2006).* Beijing, China.

Furht, B. (2008). *Encyclopedia of multimedia.* Springer.

Garbas, J., Fecker, U., Troger, T., & Kaup, A. (2006). 4D scalable multi-view video coding using disparity compensated view filtering and motion compensated temporal filtering. *Proceedings of the IEEE International Workshop on Multimedia Signal Processing 2006 (MMSP06).* Victoria, Canada.

Gokturk, S., Yalcin, H., & Bamji, C. (2005). A time-of-flight depth sensor - system description, issues and solutions. *IEEE Comp. Soc. Conf. on Comp. Vis. and Pat. Recog. Workshops 2004* (pp. 35-45). IEEE.

Guo, X., Lu, Y., Wu, F., & Gao, W. (2006). Inter-view direct mode for multiview video coding. *IEEE Trans. on Circ. and Syst. for Video Tech. , 16* (12), pp. 1527–1532.

Hartley, R., & Zisserman, A. (2003). *Multiple view geometry in computer vision.* Cambridge, U.K: Cambridge Univ. Press.

He, R., Yu, M., Yang, Y., & Jiang, G. (2008). Comparison of the depth quantification method in terms of coding and synthesizing capacity in 3DTV system. *Proc. of the 9th Intl Conf. on Sig. Proc. (ICSP) 2008*, (pp. 1279–1282). Leipzig, Germany.

Hewage, C., Karim, H., Worrall, S., Dogan, S., & Kondoz, A. (2007). Comparison of stereo video coding support in MPEG-4 MAC, H.264/AVC and H.264/SVC. *Proc. of IET Visual Inf. Eng.-VIE07*.

ITU-T Recommentation H.264 & ISO/IEC 14496-10 AVC. (2005). *Advanced video coding for generic audio-visual services, version 3*.

Javidi, B., & Okano, F. (2002). *Three-dimensional television, video, and display technologies*. Springer-Verlag.

Jebara, t. A., & Pentland, A. (1999). 3-D structure from 2-D motion. *IEEE Sig. Proc. Mag.*, *16* (3), pp. 66–84.

Kanatani, K. (1990). *Group-theoretical methods in image understanding* (Vol. 20). Springer Series in Information Sciences.

Kauff, P., Atzpadin, N., Fehn, C., Muller, M., Schreer, O., Smolic, A., et al. (2007). Depth map creation and image based rendering for advanced 3DTV services providing interoperability and scalability. *Signal Processing: Image Communication, Special Issue on 3DTV*, *22* (2), pp. 217-234.

Kienzle, W., Schölkopf, B., Wichmann, F., & Franz, M. (2007). How to find interesting locations in video: a spatiotemporal interest point detector learned from human eye movements. *Patt. Rec. (DAGM 2007)* (pp. 405-411). Darmstadt, Germany: Springer, LNCS.

Lucchese, L., Doretto, G., & Cortelazzo, G.M. (2002). A frequency domain technique for 3D view registration, *IEEE Trans. on Pat. Analysis and Machine Int.*, Vol. 24, No. 11, pp. 1468–1484.

Magnor, M., Ramanathan, P., & Girod, B. (2003). Multi-view coding for image-based rendering using 3-D scene geometry. *IEEE Trans. Circ. and Systems for Video Techn.*, *13* (11), pp. 1092–1106.

Manjunath, B., Salembier, P., & Sikora, T. (2002). *Introduction to MPEG-7: multimedia content description language*. John Wiley & Sons.

Martinian, E., Behrens, A., Xin, J., Vetro, A., & Sun, H. (2006). Extensions of H.264/AVC for multiview video compression. *Proc. of IEEE Intl Conf. on Image Proc. (ICIP2006)*. Atlanta, GA, USA.

Micallef, B. W., Debono, C., & Farrugia, R. (2010). Exploiting depth information for fast multi-view video coding. *IEEE Proc. of Int. Picture Coding Symposium 2010 (PCS 2010)*. Nagoya, Japan.

Micallef, B., Debono, C., & Farrugia, R. (Sep. 2010). Error concealment techniques for H.264/MVC encoded sequences. *IEEE Proc. of Int. Conf. of Electrotec. and Comp. Sc. (ERK)*. Portoroz, Slovenia.

Morvan, Y., Farin, D. & de With, Peter H. N. (2006). Design considerations for view interpolation in a 3D video coding framework, *27th Symp. on Inf. Theory in the Benelux*, vol. 1, Noordwijk, The Netherlands.

Oliensis, J. (2000, Nov.). A critique of structure from motion algorithms. *Comput. Vis. Image Underst.*, *80* (2), pp. 172–214.

Ozaktas, H., & Onural, L. (2008). *Three-dimensional television capture, transmission, display*. Springer-Verlag.

Pourazad, M., Nasiopoulos, P., & Ward, R. (2009, May). An H.264-based scheme for 2D to 3D video conversion. *IEEE Trans. on Consumer Electronics*, *55* (2), pp. 742-748.

Richardson, E., & et al., e. (2003). *H.264 and MPEG-4 video compression: video coding for next-generation multimedia*. Chichester: John Wiley & Sons Ltd.

Schmid, C., Mohr, R., & Bauckhage, C. (2000). Evaluation of interest point detectors. *Int. J. Comput. Vis.*, *37* (2), pp. 151–172.

Schuur, K., Fehn, C., Kauff, P., & Smolic, A. (2002). *About the impact of disparity coding on novel view synthesis, MPEG02/M8676 doc*. Klagenfurt.

Schwarz, H., Marpe, D., & Wiegand, T. (2006). Analysis of hierarchical B pictures and MCTF. *Proc. of IEEE Intl Conf. on Multimedia and Expo (ICME 2006)*. Toronto, Ontario, CA.

Sikora, T. (2001). The MPEG-7 visual standard for content description – an overview. *IEEE Trans. on Circuits and Syst. for Video Tech.*, *11* (6), pp. 696–702.

Smolic, A., & McCutchen, D. (2004). 3DAV exploration of video-based rendering technology in MPEG. *IEEE Trans. on Circuits and Syst. for Video Tech. , 14* (3), pp. 348–356.

Sorbier, F. d., Takaya, Y., Uematsu, Y., Daribo, I., & Saito, H. (Oct. 2010). Augmented reality for 3D TV using depth camera input. *IEEE VSMM 2010.* Seoul, Korea.

Starck, J., Kilner, J., & Hilton, A. (2008). Objective quality assessment in free-viewpoint video production. *IEEE Proc. of 3DTV08*, (pp. 225–228).

Stewart, J., Yu, J., Gortler, S., & Mcmillan, L. (2003). A new reconstruction filter for undersampled light fields. *Eurographics Symposium on Rendering, ACM Intl Conf. Proceeding Series*, (pp. 150–156).

Stoykova, E., Alatan, A., Benzie, P., Grammalidis, N., Malassiotis, S., Ostermann, J., et al. (2007). 3D time varying scene capture technologies - a survey. *IEEE Trans. on Circ. and Syst. for Video Tech. , 17* (11), pp. 1568-1586.

Tanimoto, M., & Fuji, T. (2003). Ray-space coding using temporal and spatial predictions. *ISO/IEC JTC1/SC29/WG11 Document M10410* .

Vetro, A., Su, Y., Kimata, H., & Smolic, A. (2006). Joint multi-view video model, Joint Video Team, *Doc. JVT-U207*. Hangzhou, China.

Vlasic, D., Peers, P., Baran, I., Debevec, P., Popovic, J., Rusinkiewicz, S., et al. (2009). Dynamic shape capture using multi-view photometric stereo. *ACM Transactions on Graphics , 28* (5).

Wang, Z., & Bovik, A. (2006). *Image quality assessment.* New York, NY, USA.

Yang, J., Everett, M., Buehler, C., & Mcmillan, L. (2002). A real-time distributed light field camera. *Proc. of the 13th Eurographics Workshop on Rendering, Eurographics Association*, (pp. 77–86).

Zitnick, C., Kang, S., Uyttendaele, M., Winder, S., & Szeliski, R. (2004, August). High-quality video view interpolation using a layered representation. *ACM Transactions on Graphics (TOG) , 23* (3).

**ADDITIONAL READING SECTION**

Balasubramanian, R., Das, S. & Swaminathan, K. (2003). Simulation studies for the performance analysis of the reconstruction of a line in 3-D from two arbitrary perspective views using two plane intersection method, *Intl J. of Computer Mathematics*, Vol. 80, No.5, pp. 559-571.

Basha, T., Moses, Y. & Kiryati, N. (2010). Multi-view scene flow estimation: a view centered variational approach. *Proc. of CVPR, 2010*.

Ben-Ari, R. & Sochen, N. (2009). A geometric framework and a new criterion in optical flow modeling. *Journal of Mathematical Imaging and Vision*, 33(2):178–194.

Chiuso, A. &. Soatto, S. (2000). Motion and structure from 2-D motion causally integrated over time: Analysis. *IEEE Transactions on Robotics and Automation* , 24:523–535.

Cyganek, B. & Siebert,J.P.(2009). *An Introduction to 3D Computer Vision Techniques and Algorithms*, John Wiley& Sons.

Dahmen, H., Franz, M.O., & Krapp, H. (2001). Extracting egomotion from optic flow: limits of accuracy and neural matched filters. In J. Zanker, & J. Zeil (Eds.), *Motion Vision: Computational, Neural and Ecological Constraints* (pp. 143-168). Berlin: Springer.

Daribo, I., Kaaniche, M., Miled, W., Cagnazzo, M., & Pesquet-Popescu, B. (2009). Dense disparity estimation in multiview video coding. *IEEE MMSP09.* Rio de Janeiro, RJ, Brazil.

Favaro, P. & Soatto, S. (2007). *3-D Shape Estimation and image restoration-exploiting defocus and motion blur*. Springer-Verlag.

ISO/IEC JTC 1. ( 2001). *Coding of audio-visual objects – Part 2: Visual.* ISO/IEC 14496-2 (MPEG-4 visual version 1), April 1999; Amd. 1 (ver. 2), February, 2000; Amd. 2, 2001, Amd. 3, 2001, Amd. 4 (streaming video profile), Amd 1 to 2nd ed. (studio profile).

ITU-T and ISO/IEC JTC 1 . (1994). *Generic coding of moving pictures and associated audio information – part 2: Video.* Recommendation H.222.0 and ISO/IEC 13 818-2 (MPEG-2 Video).

Jurie, F., & Dhome, M. (2002, February). Real time tracking of 3D objects: an efficient and robust approach. *Pattern Recognition , 35* (2), pp. 317-328.

Kumar, S., Sukavanam, N. & Balasubramanian, R. (2006) Reconstruction of quadratic curves in 3-d from two or more arbitrary perspective views: simulation studies, *Proc. of the SPIE Vision Geometry XIV, IS&T/SPIE Int.l Symp. on Electronic Imaging - 2006*, Vol. 6066, pp. 6066M1-6066M11 (180-190), San Jose, California, USA.

Levin, A., Fergus, R., Durand, F. & Freeman, W. T. (2007). Image and depth from a conventional camera with a coded aperture. Proc. SIGGRAPH, ACM.

Ma, Y., Soatto, S., Kosecka, J. & Sastry, S.S. (2003). *An Invitation to 3-D Vision*, Springer

Martens, J. (2002). Multidimensional modeling of image quality. *Proc. of the IEEE , 90*, pp. 133–153.

*Matusik, W., & Pfister, H. (2004). 3D TV: a scalable system for real-time acquisition, transmission and autostereoscopic display of dynamic scenes. 23 (3), pp. 814--824.*

Neumann, J., Fermuller, C. & Aloimonos, Y. (2003). Polydioptric camera design and 3d motion estimation. *Proc. of CVPR*, 2003.

Smolic, A. & Kauff, P. (2005). Interactive 3D video representation and coding technologies, *Proceedings of the IEEE*, vol. 93, no. 1, pp. 98–110.

Tam, W., Soung Yee, A., Ferreira, J., Tariq, S., & Speranza, F. (2005). Stereoscopic image rendering based on depth maps created from blur and edge information. *Stereoscopic Displays and Applications XII , 5664*, pp. 104-115.

Tam, W., Speranza, F., Zhang, L., Renaud, R., Chan, J., & Vazquez, C. (2005). Depth image based rendering for multiview stereoscopic displays: role of information at object boundaries. *Three-Dimensional TV, Video, and Display IV , 6016*, pp. 75-85.

Vedula, S., Rander, P., Collins, R. & Kanade, T. (2003). Three-dimensional scene flow. *PAMI*.

VQEG. (Mar. 2000). *Final report from the video quality experts group on the validation of objective models of video quality assessment. Phase I.* http://www.vqeg.org/.

VQEG. (2003). *Final report from the video quality experts group on the validation of objective models of video quality assessment. Phase II.* http://www.vqeg.org/.

Wang, Z., & Bovik, A. (2009). Mean Squared Error: Love it or Leave it. *IEEE SP Mag , 98*.

Wildeboer, M., Fukushima, N., Yendo, T., Panahpour Tehrani, M., Fujii, T., & Tanimoto, M. (2010). A semi-automatic multi-view depth estimation method. (P. Frossard, H. Li, F. Wu, B. Girod, S. Li, & G. Wei, Eds.) *Proc. of the SPIE, Visual Comm. and Image Proc. , 7744*, pp. 77442B-77442B-8.

Wöhler, C. (2009). *3D Computer Vision: Efficient Methods and Applications*, Springer Verlag.

Zhu, C., Zhao, Y., Yu, L. & Tanimoto, M.(2012) *3D-TV System with Depth-Image-Based Rendering: Architectures, Techniques and Challenges,* Springer Verlag.

**KEY TERMS & DEFINITIONS**

Keywords: 3D TV, motion estimation, motion compensation, video streaming, multimedia, 4D imaging, image processing, video coding, video compression, error concealment.